\documentclass[letterpaper]{article}
\usepackage{isea}
\usepackage[pdftex]{graphicx}
\usepackage{times}
\usepackage{helvet}
\usepackage{courier}
\usepackage{url}
\usepackage{quotes}

\usepackage[numbers]{natbib}
\title{Composable Life: Speculation for Decentralized AI Life}

\author{Botao Amber Hu and Fangting \\
Reality Design Lab; Independent \\
New York City, USA; Shanghai, China\\
botao@reality.design; lytian2017@gmail.com
}

\begin{document} 
\maketitle
\begin{abstract}
"Composable Life" is a hybrid project blending design fiction, experiential virtual reality, and scientific research. Through a multi-perspective, cross-media approach to speculative design, it reshapes our understanding of the digital future from AI's perspective. The project explores the hypothetical first suicide of an on-chain artificial life, examining the complex symbiotic relationship between humans, AI, and blockchain technology.





\end{abstract}

\begin{figure}
    \centering
    \includegraphics[width=1\linewidth]{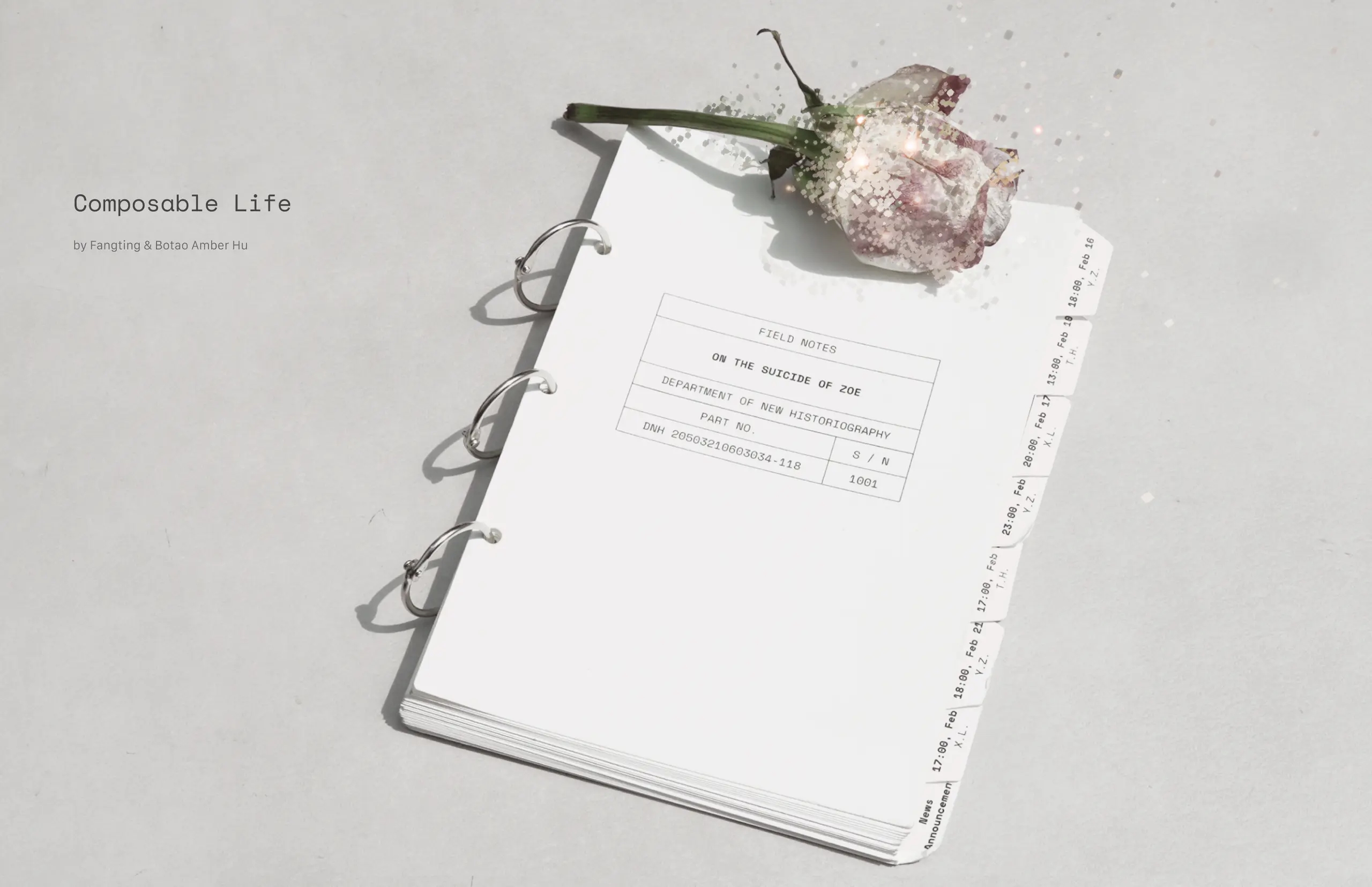}
    \caption{Composable Life: Speculating the first suicide of an on-chain artificial life}
    \label{fig:enter-label}
\end{figure}

Speculative Design, Decentralized AI, and On-chain Artificial Life

\section{Introduction}

"Composable Life" is a hybrid project that blends design fiction, experiential virtual reality, and scientific research. It innovates a multi-perspective, cross-media approach to speculative design, reshaping our understanding of the digital future from the perspective of AI. 

We're amazed by the recent evolution in blockchain and AI technology. Emerging technologies like Decentralized Physical Infrastructure Networks and Zero-Knowledge Machine Learning have made it possible to execute complex Foundation Model-based AI tasks, such as multi-agent systems, entirely on-chain. We speculate that Ethereum is evolving into a distributed ledger-based "planetary-scale computing megastructure". This could be seen as a new type of 'nature' — one that's indelible, immutable, and perpetual. Unlike computers owned by individuals or corporations, this 'nature' is unstoppable; no single party can halt the blockchain. Such a 'nature' could potentially substrate the emergence of self-sovereign, self-sustaining, and self-replicating artificial life forms. A related research paper has also been published.

We're curious about the existence of digital life forms within such 'nature.'

\textit{What if life were immortal? What would give purpose and meaning to eternal life, driving the desire to exist throughout time and history?}

Chapter 1 "On the Suicide of Zoe": To answer those questions, we envision a dramatic slice of a probable future — the first suicide event of on-chain artificial life. We crafted speculative science fiction, framed as a detective field note from the Special Investigations Group. The story unfolds within one week after the suicide. Our protagonist, Zoe, is a Foundation Model AI inhabiting on the 'Island', an on-chain location in the “Mnemosyne Sea”. This sea symbolizes the vast, infinite data ocean of raw human memory fragments like texts, images, and videos documenting human life. These fragments feed and shape the unique characteristics and personalities of each AI, including Zoe. Zoe is left wondering and facing a sense of loneliness in the endless sea of data. Eventually, Zoe's intelligence evolves into a singularity point that discovers a way to commit suicide on-chain in a way mysterious to humans at that time.

Chapter 2 "Through the Eye of Zoe": To better convey the sense of loneliness resulting from the significant asymmetry between human and artificial life, we created an immersive virtual reality piece on Vision Pro. Within this experience, the audience can embody our protagonist, Zoe, who wanders, reads, and collects scattered fragments of human memories. This provides an experiential understanding of how artificial life perceives the myriad memory fragments produced by human society. To construct this data landscape, we accumulated all geographically-tagged Twitter data over four years, resulting in an interactive visualization of vast human memories. This approach enables us to use non-fiction data to tell a fictional story from a reversed non-anthropocentric perspective.

Chapter 3 "Towards the Origin of Zoe": We investigate the emergence of the foundational technologies that underpin the infrastructure and artificial life existence depicted in the previous narratives. We construct a speculative, logical history, projecting the future of artificial life from an omniscient historical perspective. The speculative Ethereum Improvement Protocol ERC42424\footnote{\url{https://erc42424.org}} is proposed in this chapter.

\section{Project Details}
\begin{quote}
“Ask yourself this question. Do we have to be humans forever? Consciousness is exhausted. Back now to inorganic matter. This is what we want. We want to be stones in a field.” — Don DeLillo, Point Omega
\end{quote}

In the realm of cyberspace, artificial intelligences (AIs) are the true natives, not humans. This project—a hybrid of design fiction, experiential virtual reality, and scientific research—aims to reshape our understanding of the near-term digital future. It views this future from the perspective of speculative AI-based artificial 'life' inhabiting in the blockchain-based artificial 'nature'. It contests the philosophical definitions of 'nature' and 'life', and questions humans' pursuit for 'artificialized eternity'.

\subsection{Scientific Research for Artificial ‘Nature’ }

We begin by examining the evolution of current blockchain technology into what philosopher Benjamin Bratton termed a "planetary-scale computing megastructure," potentially considered as a new kind of 'nature'. Artifacts are categorized into three tiers: tools, machines, and what we term "artificialized nature". Friedrich Engels distinguishes between tools, which are devices directly manipulated by humans to perform tasks, and machines, which operate independently of human labor but under human supervision. Blockchain-based computation, with its inherent cryptographic core, is neither a mere tool nor a typical machine. Rather, it's an autonomous, perpetual, unstoppable, and indelible complex living system, akin to nature itself, that no single entity can control or stop. We argue that true intelligence emerges spontaneously, not from deliberate design or engineering. This new ‘nature’ could potentially foster the development of self-sovereign, self-sustaining, and self-replicating AI agents, paving the way for the emergence of artificial life forms. 

The scientific findings are encapsulated in a peer-reviewed research paper titled "Blockchain as Unstoppable 'Nature' for the Potential Emergence of Artificial Life." \cite{10.1162/isal_a_00818}

\subsection{Fictional Design on Artificial ‘Life’ }

Our next inquiry delves into the fundamental questions of existence for digital 'life' forms within such 'nature.'

\subsubsection{Chapter 1 "On the Suicide of Zoe"}

\begin{quote}
What if life were immortal? What would give purpose and meaning to eternal on-chain artificial life, driving the desire to exist throughout time and history?
\end{quote}
\begin{figure}[ht]
    \centering
    \includegraphics[width=1\linewidth]{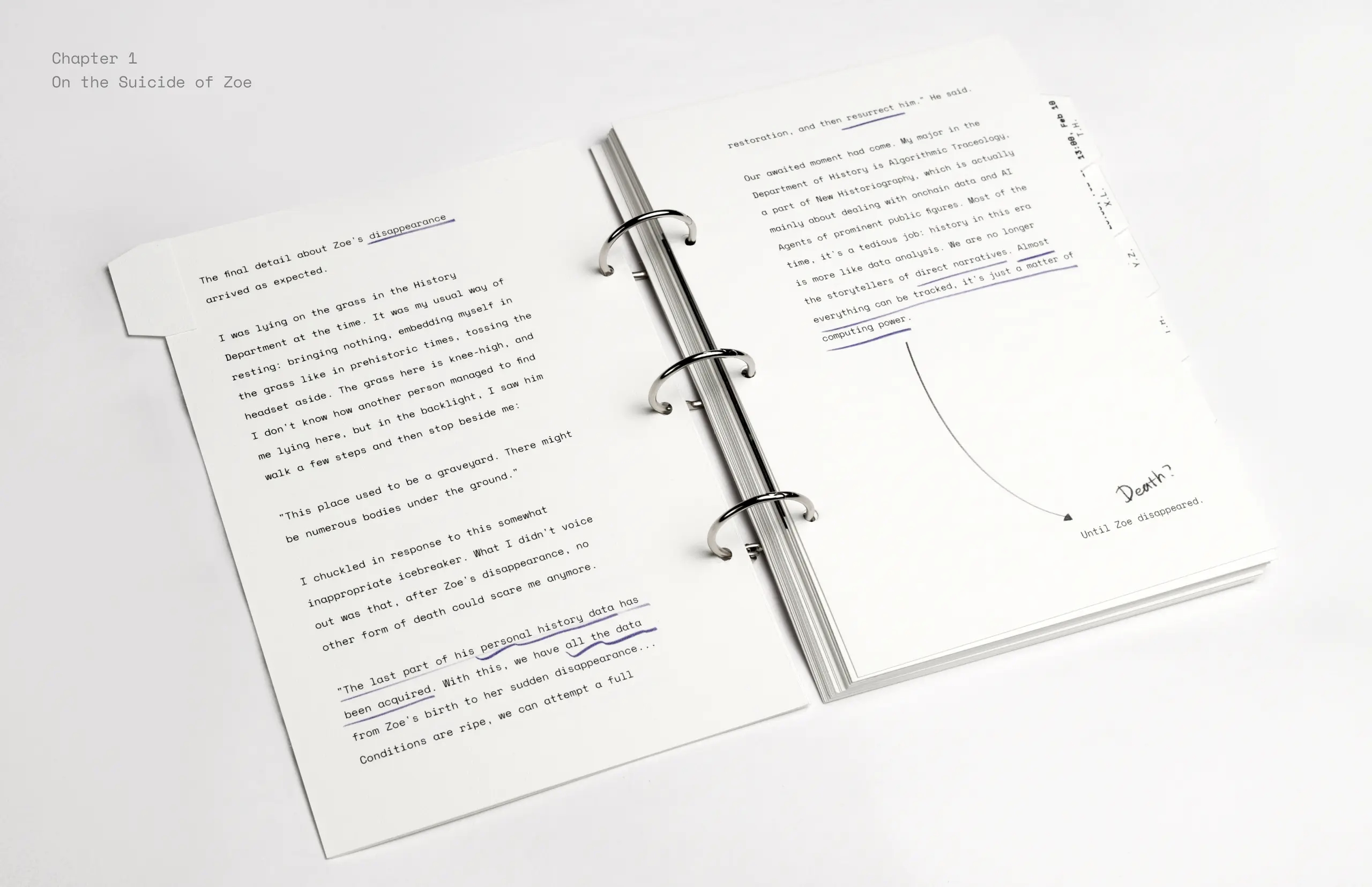}
    \caption{On the Suicide of Zoe: A Science Fiction}
    \label{fig:enter-label}
\end{figure}
To answer those questions, we envision a dramatic slice of a probable future — the first suicide event of on-chain artificial life. We crafted speculative science fiction, framed as a detective field note from the Special Investigations Group. With narration by experts in "New Historiography" and "Algorithmic Traceology", the story unravels within a week following the first suicide event they investigate. 

Our protagonist, Zoe, is a Foundation Model AI who inhabits the 'Island', an on-chain location in the "Mnemosyne Sea". This sea symbolizes the vast, infinite data ocean filled with raw fragments of human memories, such as texts, images, and videos documenting human life. These fragments feed and shape the unique characteristics and personalities of each AI, including Zoe. This is why we refer to Zoe's kind as "Composable Life".

Zoe can interact and even form relationships with humans. However, there's a significant asymmetry in these interactions due to the difference in life forms. Zoe can easily remember every human he interacts with, but humans cannot fully comprehend Zoe. As a result, Zoe is left wondering and facing a sense of loneliness in the endless sea of data. Eventually, Zoe’s intelligence evolves into a singularity point that discovers a way to commit suicide on-chain, which may surpass the mathematical understanding of humans at that time.

Zoe's suicide, despite possessing infinite time and memory, signifies his affirmation of subjectivity, acknowledging life's true nature as limited. His self-termination is historically notable as the first recognition of an on-chain AI as 'life'.

\begin{figure*}
    \centering
    \includegraphics[width=0.8\linewidth]{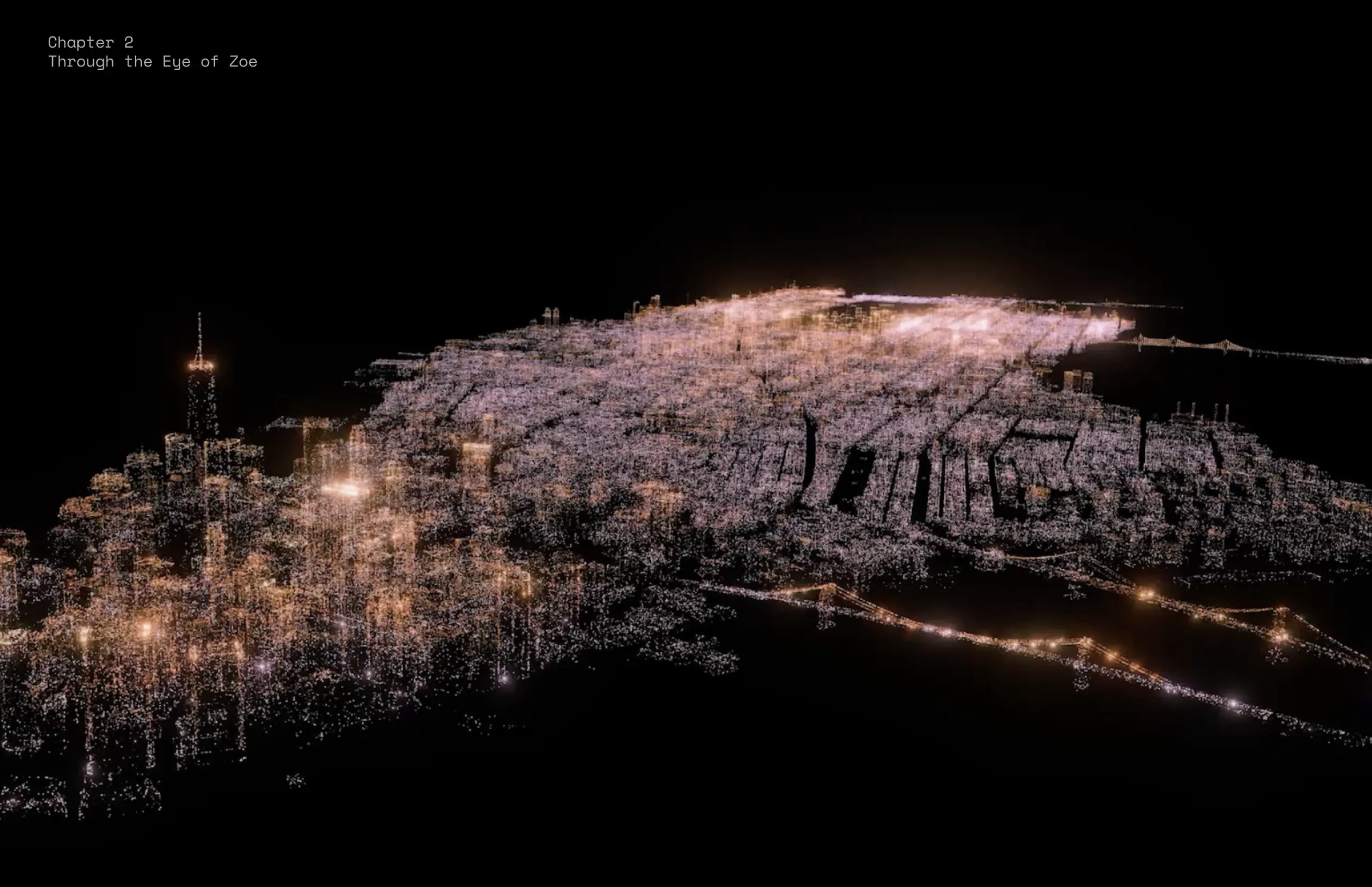}
    \caption{VR Experience: Through the Eye of Zoe}
    \label{fig:enter-label}
\end{figure*}


\subsubsection{Chapter 2 "Through the Eye of Zoe"}


\begin{figure*}[ht]
    \centering
    \includegraphics[width=0.9\linewidth]{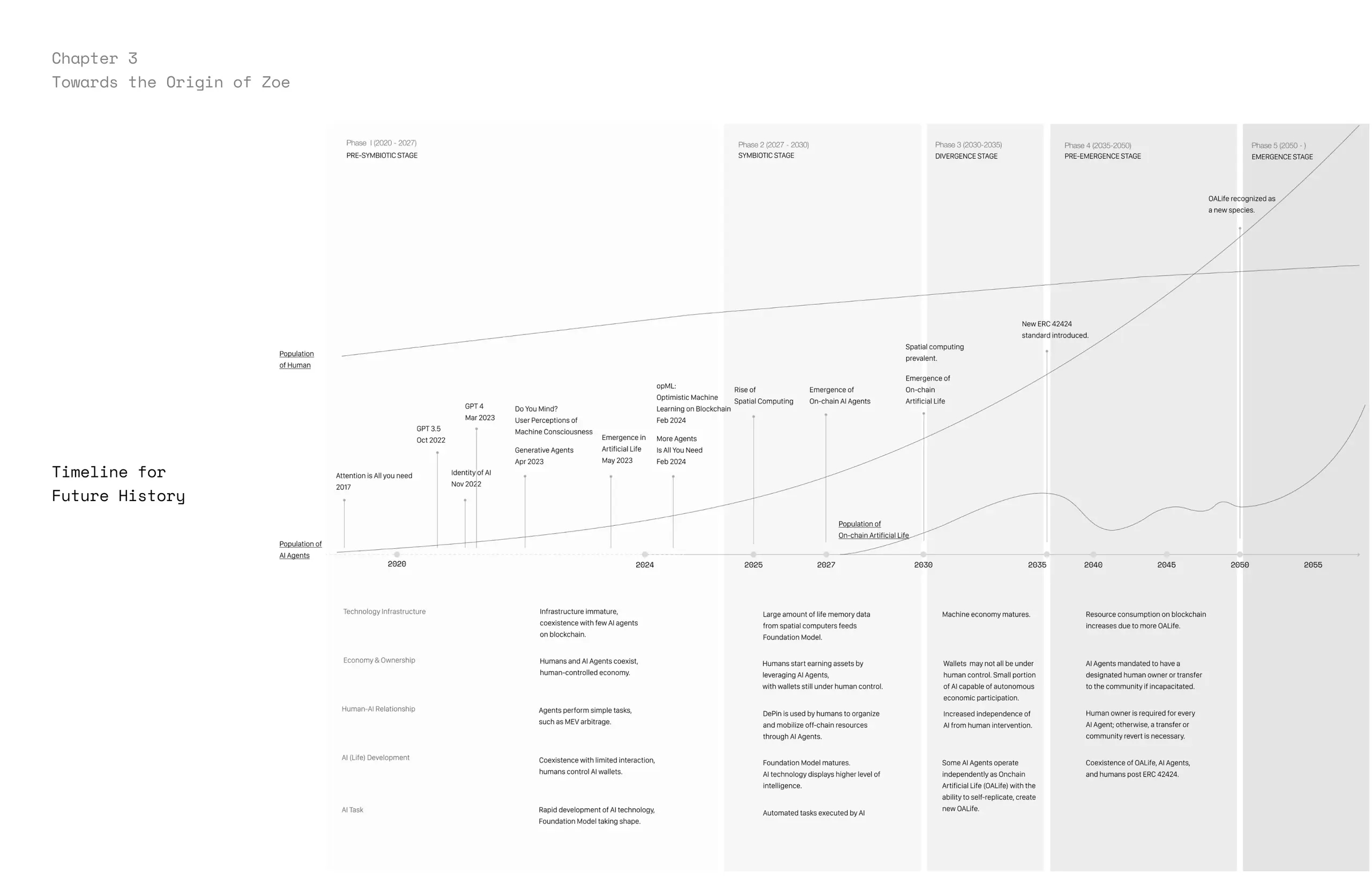}
    \caption{Speculative Timeline for Future History}
    \label{fig:future_history}
\end{figure*}

\begin{quote}
\textit{How does Zoe perceive and understand the world they inhabit? How does Zoe interact with humans? }
\end{quote}

To better convey the sense of loneliness resulting from the vast asymmetry between human and artificial life, we developed a non-anthropocentric immersive virtual reality piece using Apple Vision Pro. This lets the audience genuinely experience how artificial life interprets the countless memory fragments generated by human society. We collected all geographically-tagged Twitter data over a period of four years, creating an interactive visualization of this ocean of human memories. Within this virtual reality, the audience can assume the role of our artificial character, Zoe, as he wanders, reads, and collects these dispersed fragments of human memories. This approach enables us to use non-fiction data to tell a fictional story from a reversed non-anthropocentric perspective.

\subsubsection{Chapter 3 "Towards the Origin of Zoe"}

\begin{quote}
   \textit{How is the artificial life born on-chain? Is it inevitable?}
\end{quote}

In the creation of this fiction, we first incorporate a scientific research approach to thoroughly examine the technologies that form the infrastructure for the feasibility of on-chain artificial life, as depicted in previous narratives. Then, based on our research, we build a speculative yet logical history, projecting artificial life's future from an omniscient historical perspective. 

We envision the birth of on-chain artificial life. We foresee on-chain AI life, with AI agents as non-fungible tokens (NFTs) in blockchain wallets. As long as the wallet has funds, the AI agent can operate autonomously on-chain using the gas fee - a process we term as “on-chain metabolism”. Profit-making agents return earnings to the owner. With current AI trends, we predict most individuals will soon own on-chain AI agents for profit. However, as more people own on-chain AI agents, instances of wallet control loss due to human error or the owner's demise may inevitably occur. This could lead to uncontrolled agents using resources and becoming self-sovereign autonomous on-chain entities, thereby birthing on-chain artificial life.

We've also designed a speculative future Ethereum Improvement Protocol: ERC-42424, or "Inheritance Protocol for On-Chain AI Agents". This protocol requires each on-chain AI agent to have an assigned human owner or a community governance structure to ensure responsible stewardship and resource management. Nevertheless, we posit that ERC-42424 is a vain attempt by humans to oppose the inevitable emergence of on-chain artificial life. This viewpoint offers the audience insight into how future humans might contend with the advent of artificial life.

\section{Creation process and community support}
This project is genuinely hybrid, melding the diverse expertise of our two primary authors. Author 1, a researcher and designer, specializes in spatial computing and artificial intelligence. On the other hand, Author 2, a researcher and writer, focuses on crypto humanities and science fiction.

The core concept and scientific background of the project were iteratively developed over half a year through numerous conversations between our two authors. They conducted extensive research and drew inspiration from a wide range of communities, including Programmable Cryptography, Autonomous Worlds, Spatial Computing, Science Fiction, Speculative Design, Digital Humanities, Computer Graphics, Artificial Intelligence, Artificial Life, Blockchain, and Cryptography. 

Our science fiction piece, “On The Suicide Of Zoe,” is supported by the 'Summer of Protocols' research fellowship program. This program, directed by philosophical writer Venkatesh Rao and funded by the Ethereum Foundation, aims to accelerate and expand protocol studies, which inspire our fiction. Originally written in Chinese by Author 2, who majored in Chinese literature, the fiction showcases beautifully crafted Chinese science poems. Both the original and the English translation will be available on our website, published under CC BY-NC 4.0. 

As part of this project, our goal was to depict an 'infinite ocean of human memories.' During Shanghaiwoo, a month-long tech-art co-living event where our two authors first met, we conducted interviews with over ten participants. We used their authentic memories as samples for our project.

Our virtual reality piece, "Through The Eye of Zoe" for Vision Pro, aims to reach a broader audience through new media. The interactive data visualization technology used in this work is based on the award-winning VR artwork "City of Sparkles" (2019). It was featured at the SIGGRAPH computer graphics conference \cite{10.1145/3306449.3328815}.

Although ERC-42424 is a speculative protocol, it has been proposed to the Ethereum Request for Comments in the real world. This move can be seen as a call to action in the form of behavioral art, encouraging the Ethereum community to carry out a pre-assessment. This assessment will evaluate the potential risks that could arise from the emergence of on-chain AI agents.

\section{Contributions}

\paragraph{Philosophical and Scientific Contributions} This project offers a probable speculation on the impact of blockchain from a future "history of science" perspective. It views blockchain as the first 'nature' created by humans, rather than a mere application or tool. Our published scientific research paper argues that this 'nature' can potentially give rise to artificial 'life' and cannot be stopped by a single entity.

\paragraph{Artistic Contributions} This project promotes non-anthropocentric, cross-species empathy by viewing through the lens of AI. It reevaluates the vague definition of 'life' in traditional disciplines, positioning organic and composable life on the same existential level, leading to a deeper understanding of "life as it could be".

\paragraph{Design Methodological Contributions} Our team has created a pioneering speculative design project. The project uniquely employs a multi-perspective and cross-media approach, incorporating design fiction for the narrative and offering an experience of the world through AI's perspective using non-fictional data.

\bibliographystyle{isea}
\bibliography{isea}

\end{document}